\documentclass[prd,showpacs,preprint,nofootinbib]{revtex4-1}
\usepackage{amsmath,amssymb}
\usepackage[colorlinks=true,linkcolor=blue]{hyperref}

\newcommand{\nn}{\nonumber}
\def\/{\over}
\newcommand{\bra}[1]{\langle#1|}
\newcommand{\ket}[1]{|#1\rangle}

\begin{document}

\title{Lightcone fluctuations in a nonlinear medium due to thermal fluctuations}

\author{Jiawei Hu$^{1}$ and Hongwei Yu$^{1}\footnote{Corresponding author at hwyu@hunnu.edu.cn}$}
\affiliation{
$^1$ Department of Physics and Synergetic Innovation Center for Quantum Effects and Applications, Hunan Normal University, Changsha, Hunan 410081, China}

\begin{abstract}

We study the flight time fluctuations of a probe light propagating in a slab of nonlinear 
optical material with an effective fluctuating refractive index caused by thermal fluctuations of background photons at a temperature $T$, which are analogous to the lightcone fluctuations due to fluctuating spacetime geometry when gravity is quantized. A smoothly varying second order susceptibility is introduced, which results in that background field modes whose wavelengths are of the order of the thickness of the slab give the main contribution. We show that, in the low-temperature limit, the contribution of thermal fluctuations to the flight time fluctuations is proportional to $T^4$, which is a small correction compared with the contributions from vacuum fluctuations, while in the high-temperature limit, the contribution of thermal fluctuations increases linearly with $T$, which dominates over that of vacuum fluctuations.  Numerical estimation shows that, in realistic situations, the contributions from thermal fluctuations are still small compared with that from vacuum fluctuations even at room temperature.

\end{abstract}

\maketitle

\section{Introduction}

As necessitated by the uncertainty principle, lightcones are expected to be fluctuating in any theory of quantum gravity. It was first conjectured by Pauli that the ultraviolet divergences in quantum field theory, which arise from the lightcone singularities of two-point functions, might be removed when gravity is quantized, since lightcones are supposed to be smeared out due to the fluctuations of the spacetime metric \cite{Pauli}. This idea was further investigated by several authors  \cite{Deser,DeWitt,Isham71,Isham72}. A direct consequence of lightcone fluctuations is that the flight time of a probe light from its source to a detector is no longer fixed, but undergoes fluctuations around its classical value \cite{Ford95,Ford96,Yu99,Ellis,Yu00,Yu03,Yu09,Mota}.

In principle, the flight time fluctuations are observable. However, this effect is generally too small to be observed. Recently, Ford {\it et al}. proposed that a nonlinear medium with a fluctuating background field may be considered as an analogue system for quantum lightcone fluctuations \cite{analog1,analog2}. In a nonlinear medium, the effective refractive index for a probe light is fluctuating when the medium is subjected to a fluctuating background field, which leads to a fluctuating flight time. This is in close analogy to the lightcone fluctuation due to metric fluctuations when gravity is quantized. Besides the analogue models for active gravitational field fluctuations, which are analogous to the fluctuations of the dynamical degrees of freedom of gravity itself \cite{analog1,analog2}, there are also analogue models for passive fluctuations of gravity driven by the fluctuations of matter fields \cite{analog3,analog4}. The fluctuating background field can be either a squeezed vacuum \cite{analog1,analog3}, or a bath of fluctuating electromagnetic fields in vacuum \cite{analog2,analog4}. Apart from being an analogue model for quantum gravity, the work  \cite{analog2,analog4} also provides a model to show that vacuum fluctuations sampled in a finite time scale are potentially observable. Therefore, a natural question is how this effect may be affected by thermal fluctuations which are unavoidable in an actual experiment. In this paper, we study the contribution of thermal fluctuations to the flight time fluctuations in detail. In particular, we are interested in whether we need a very low temperature environment if we aim at observing the lightcone fluctuations in a nonlinear medium due to vacuum fluctuations.  The paper is organized as follows: In Sec. \ref{sec2}, we briefly introduce the basic formalism of quantum lightcone fluctuations in a nonlinear medium \cite{analog1,analog2}. In Secs. \ref{sec3}, we investigated the contribution of thermal fluctuations to the lightcone fluctuations. Finally, we give a summary in Sec. \ref{sec4}. The Lorentz-Heaviside units with $\hbar=c=k_B=1$ are used in this paper unless specified.

\section{The Basic Formalism} \label{sec2}

To begin, we review the lightcone fluctuations in quantum gravity. Consider a flat spacetime $\eta_{\mu\nu}$ with a linear perturbation $h_{\mu\nu}$, so the spacetime metric $g_{\mu\nu}$ can be written as
\begin{equation}
g_{\mu\nu}=\eta_{\mu\nu}+h_{\mu\nu}.
\end{equation}
Define $\sigma$ as one-half of the squared geodesic distance between two spacetime points $x$ and $x'$, so it can be expanded with respect to $h_{\mu\nu}$ as
\begin{equation}
\sigma=\sigma_0+\sigma_1+O(h_{\mu\nu}^2).
\end{equation}
If $h_{\mu\nu}$ is quantized, the lightcone is expected to be fluctuating as a result of the quantum gravitational vacuum fluctuations. In particular, we have $\langle\sigma_1^2\rangle\neq0$. Such quantum gravitational effect is in principle observable by considering the propagation of light pulses from their source to a detector separated by a distance $r$. The root-mean-square deviation of the flight time is \cite{Ford95}
\begin{equation}
\Delta t=\frac{\sqrt{\langle\sigma_1^2\rangle}}{r}.
\end{equation}

In a nonlinear medium, the fluctuations of background field will cause a fluctuating effective refractive index for a probe light propagating in it, which leads to a fluctuating flight time \cite{analog1,analog2}. The source free Maxwell's equations in a dielectric medium can be written as 
\begin{eqnarray}
\nabla\cdot\mathbf{D}=0,\qquad
\nabla\times\mathbf{E}=-\frac{\partial\mathbf{B}}{\partial t},\qquad
\nabla\cdot\mathbf{B}=0,\qquad
\nabla\times\mathbf{H}=\frac{\partial\mathbf{D}}{\partial t}.
\end{eqnarray}
Here $\mathbf{E}$ and $\mathbf{B}$ are the electric and the magnetic fields respectively, and  $\mathbf{D}$ and $\mathbf{H}$ are the corresponding induced fields. The constitutive relations are $\mathbf{B}=\mu\mathbf{H}$, and $\mathbf{D}=\varepsilon_0\mathbf{E}+\mathbf{P}$. 
In a nonlinear medium, the relation between the electric polarization $P_i$ and the electric field $E_i$ takes the form
\begin{equation}
P_i = P_i^{(1)} + P_i^{(2)}+\cdots
    = \chi_{ij}^{(1)} E^j + \chi_{ijk}^{(2)} E^j E^k + \cdots  \;,
\end{equation}
where $\chi^{(i)}$ is the {\it i}th order susceptibility tensor. Hereafter the Einstein convention is assumed for repeated index. The total electric field $E^i$ is taken to be the sum of a background field $E_{0}^i(\omega_0)$ and a probe field $E_{1}^i(\omega_1)$.  Here, the strength of the probe light $E_{1}^i$ is assumed to be much smaller than the background field $E_{0}^i$, while its frequency $\omega_1$ is much larger than that of the background field $\omega_0$ \cite{analog1,analog2}. The second order polarization $P_i^{(2)}$ can be written as~\cite{no-book}
\begin{equation}
P_i^{(2)}(\omega_m+\omega_n)
=\sum_{m,n=0}^1\chi_{ijk}^{(2)}(\omega_m+\omega_n)E_m^j(\omega_m)E_n^k(\omega_n)\;.
\end{equation}

We assume that in the frequency regime of the background field $\omega_0$, the second order susceptibility tensor $\chi^{(2)}_{ijk}(2\omega_0)$ can be neglected. So, when the probe field is absent, the relation between the electric polarization $P_i$ and the electric field $E_i$ is linear, so  the background field modes satisfy the following wave equation
\begin{equation}\label{wave-E0}
 \nabla^2 E_0^i - \frac{1}{v_B^2} \frac{\partial^2 E_0^i}{\partial t^2} = 0\;,
\end{equation}
where $v_B=1/\sqrt{1+\chi_B^{(1)}}=1/n_B$.  Here we have assumed that for the background field the medium is isotropic, i.e. $\chi_{ij}^{(1)}=\delta_{ij}\chi_B^{(1)}$, and in the frequency regime of the background field, dispersion can be neglected \cite{analog2}.
Note that the wave equation for the background field (\ref{wave-E0}) takes exactly the same form as that in the vacuum, except that the speed of light $c$ is replaced with $v_B$. As we will see later, this implies that the difference between the electric field two-point functions and the corresponding ones in vacuum is only an overall factor of $1/n_B^3$, and a replacement of the time $t$ with $t/n_B$.

For the probe field $E_{1}^i$,  we assume that it propagates in the $x$-direction and is polarized in the $z$-direction, i.e. $E_{1}^i = \delta^{iz} E_{1}(t,x)$. The wave equation for $E_{1}$ can be obtained by subtracting the wave equation of $E_0$ from that of the total field $E_0 + E_1$, and neglect terms with $E_1^2$ and $\dot{E_0}$ as \cite{analog2}
\begin{equation}\label{wave-Ep}
 \frac{\partial^2 E_{1}}{\partial x^2}
 - \frac{1}{v_P^2} 
 \left[1 + \frac{1}{n_P^2}\left(\chi_{zz j}^{(2)} + \chi_{zjz}^{(2)}\right) \right] \frac{\partial^2 E_{1}}{\partial t^2} = 0\;.
\end{equation}
The equation above describes a wave propagating with a space and time dependent phase velocity
\begin{equation}\label{v}
v \approx v_P \left[1- \frac{1}{2n_P^2} \left(\chi_{zzj}^{(2)} + \chi_{zjz}^{(2)}\right) E_{0}^j  \right]\;,
\end{equation}
where $\left| \frac{1}{2n_P^2} \left(\chi_{zzj}^{(2)} + \chi_{zjz}^{(2)}\right) E_{0}^j \right| \ll 1$ is assumed. Here, $v_P=1/\sqrt{1+\chi_P^{(1)}}=1/n_P$ 
is generally different from $v_B$ when dispersion is taken into account, and in this paper, we are concerned with the case  $n_P>n_B$. That is, the worldline of the probe field is inside the effective lightcone.

Therefore, the flight time of the probe light propagating through a nonlinear medium with a thickness $d$ is
\begin{equation}
t=\int_{0}^{d} \frac{dz}{v}  = n_P \int_{0}^{d} \left[1 + \frac{1}{2n_P^2}
\left(\chi_{zzj}^{(2)} + \chi_{zjz}^{(2)}\right) E_{0}^j(t,\vec{x}) \right]\, dx\;.
\end{equation}
The integration is along the trajectory of the probe pulse, i.e. $x = v_P t= t/n_P$. In this paper, we assume that the background field $E_0$ is that of a thermal bath of photons at a temperature $T=1/\beta$, where $\beta$ is the thermal photon wavelength. The thermal fluctuations of $E_0$ will cause fluctuations of the flight time $t$, and the relative flight time variance takes the form \cite{analog1,analog2}
\begin{equation}
\label{dt}
\delta ^2= \frac{\langle t^2 \rangle - \langle t \rangle^2}{\langle t \rangle^2}
= \frac{1}{4 n_P^4 d^2} \int_{0}^{d} dx \int_{0}^{d} dx' \left(\chi_{zzi}^{(2)} + \chi_{ziz}^{(2)}\right) \left(\chi_{zzj}^{(2)} + \chi_{zjz}^{(2)}\right) \langle E_{0}^i(t,\vec{x})  E_{0}^j (t',\vec{x}') \rangle_\beta\;,
\end{equation}
where $\langle\,\,\rangle_\beta$ denotes the expectation value over the thermal state, and we have assumed that $\langle E^j_{0}\rangle_\beta=0$. Here, the thermal expectation in the equation above is a summation over the contributions of electromagnetic field modes in all frequencies, while in  deriving the wave equation for the probe field Eq. (\ref{wave-Ep}) we have assumed that the frequency of the background field $\omega_0$ is much smaller than that of the probe field $\omega_1$. To fulfill this assumption, first we assume that the frequency of thermal photon $\beta^{-1}$, the frequency at which the background thermal radiation spectrum peaks, is much small compared with $\omega_1$.  Then we introduce an effective cutoff of the contributions from high frequency background modes, which can be realized by a smoothly varying second order susceptibility $\chi^{(2)}(x)$ along the path of the probe light \cite{analog2}. Let $\chi^{(2)}_i\equiv \frac{1}{2d}\int_{-\infty}^{\infty} dx\, \left(\chi^{(2)}_{zzi}(x)+\chi^{(2)}_{ziz}(x) \right)$, where $\chi^{(2)}_i$ is the averaged second order susceptibility along the $x$-axis. If we choose the profile of $\chi^{(2)}_{zzi}(x)$ as the Lorentzian function
\begin{equation}\label{lorentz}
\chi^{(2)}_{zzi}(x)= \frac{d^2}{\pi(x^2+d^2)} \chi^{(2)}_{zzi}\;,
\end{equation}
the relative flight time variance Eq. (\ref{dt}) can be reformed as 
\begin{eqnarray}
 \delta^2 \propto
\bigg\langle 0\bigg |
\int_{-\infty}^{\infty} d\omega e^{-|\omega| \tau} E_0^i(\omega)
\int_{-\infty}^{\infty} d\omega' e^{-|\omega'| \tau} E_0^j(\omega')
\bigg|0\bigg\rangle\;,
\end{eqnarray}
where $E_0^i(\omega)$ is the Fourier transform of $E_0^i(t)$, and $\tau=n_P\, d$.
It is clear that contributions from field modes whose wavelengths are shorter compared with the thickness of the medium will be effectively suppressed. Plugging Eq. (\ref{lorentz}) into Eq. (\ref{dt}), the relative flight time variance can be rewritten as
\begin{equation}\label{dt-1}
\delta ^2
= \frac{(\chi^{(2)}_0 )^2\,d^2}{\pi^2 n_P^4 } \int_{-\infty}^{\infty} dx \int_{-\infty}^{\infty} dx' \frac{1}{x^2+d^2} \frac{1}{x'^2+d^2}
 \langle E_{0}^i(t,\vec{x})  E_{0}^j (t',\vec{x}') \rangle_\beta\;.
\end{equation}

\section{Lightcone fluctuations due to thermal fluctuations} \label{sec3}

In this section, we study the lightcone fluctuations of a probe light pulse in a nonlinear medium due to thermal fluctuations. First, let us work out the thermal two-point function of the electric field in the dielectric. In an empty space, the two-point function for the four potential $A_\mu(x)$ at a finite temperature $T=\beta^{-1}$, $D_\beta^{\mu \nu} (x,x') = \langle 0| A^\mu (x)\, A^\nu (x') |0 \rangle_\beta$, can be written as an infinite imaginary-time image sum of the corresponding zero-temperature two-point function,  ${D^{\mu \nu}_{0}} (x-x')$, as \cite{qftcs}
\begin{equation} \label{vec-2p}
  D_\beta^{\mu \nu} (x,x')  
= \sum^{\infty}_{m=-\infty} D^{\mu \nu}_{0} (\mathbf{x-x'},t-t'+im\beta)\;.
\end{equation}
In the Feynman gauge, we have
\begin{equation}\label{vec-2p-0}
D_{0}^{\mu\nu}(x-x')=\frac{1}{4\pi^{2}}
 \frac{\eta^{\mu\nu}}{(t-t'-i\epsilon)^2-(x-x')^2-(y-y')^2-(z-z')^2}\;,
\end{equation}
where $\epsilon\to +0$, and $\eta={\rm diag} (1,-1,-1,-1)$. The electric field two-point function $\langle E_i(x)E_j(x')\rangle_\beta$ can then be expressed as
\begin{eqnarray} \label{em-2p}
  \langle E_{i}(x)E_{j}(x')\rangle_\beta
&=& \frac{1}{4\pi^{2}}\sum_{m=-\infty}^\infty
  \left(\delta_{ij}\partial_0\partial_0'-\partial_i\partial_j' \right)\nn\\
  &&~~\times \frac{1}{(x-x')^2+(y-y')^2+(z-z')^2-(t-t'+im\beta-i\epsilon)^2}\;,
\end{eqnarray}
where $\partial_i'$ denotes the differentiation with respect to $x_i'$. The two-point function of the electric field in a dielectric can then be derived by considering the fact that the net effect of a dielectric on the electric field two-point functions is an overall factor of $1/n_B^3$, and a replacement of the time $t$ with $t/n_B$ as \cite{Glauber,Barnett,analog2} 
\begin{equation}
 \bra{0} E_x(x)E_x(x') \ket{0}_\beta
=\frac{1}{\pi^2 n_B^3}\sum_{m=-\infty}^\infty
 \frac{1}{\left[\,(\Delta t+im\beta)^2/n_B^2-(\Delta x)^2\right]^{2}}\;,
\end{equation}
\begin{equation}
 \bra{0} E_y(x)E_y(x') \ket{0}_\beta = \bra{0} E_z(x)E_z(x') \ket{0}_\beta
=\frac{1}{\pi^2 n_B^3}\sum_{m=-\infty}^\infty
 \frac{(\Delta x)^2+(\Delta t+im\beta)^2/n_B^2}
 {\left[\,(\Delta t+im\beta)^2/n_B^2-(\Delta x)^2\right]^{3}}\;,
\end{equation}
where we have taken the spatial separation to be in the $x$-axis, i.e. $\Delta y=\Delta z=0$. Here the $m=0$ term corresponds to the contribution from vacuum fluctuations, which will be omitted in the following calculations as we focus on the lightcone fluctuations due to thermal fluctuations.

The relative flight time variance due to thermal fluctuations $\delta_T^2$ can then be expressed as
\begin{eqnarray}\label{dt2}
\delta_T ^2
&=&\frac{(\chi^{(2)}_x )^2\,d^2}{\pi^4 n_B^3 n_P^4 } 
   \int_{-\infty}^{\infty} dx \int_{-\infty}^{\infty} dx' 
   \frac{1}{x^2+d^2} \frac{1}{x'^2+d^2}
   {\sum_{m=-\infty}^\infty}{'}
   \frac{1}{\left[\,(\Delta t+im\beta)^2/n_B^2-(\Delta x)^2\right]^{2}}  \nn\\
&&~+\frac{(\chi^{(2)}_y )^2\, d^2}{\pi^4 n_B^3 n_P^4 } 
   \int_{-\infty}^{\infty} dx \int_{-\infty}^{\infty} dx' 
   \frac{1}{x^2+d^2} \frac{1}{x'^2+d^2}
   {\sum_{m=-\infty}^\infty}{'}
   \frac{(\Delta x)^2+(\Delta t+im\beta)^2/n_B^2}
 {\left[\,(\Delta t+im\beta)^2/n_B^2-(\Delta x)^2\right]^{3}}\nn\\
&&~+\frac{(\chi^{(2)}_z )^2\, d^2}{\pi^4 n_B^3 n_P^4 } 
   \int_{-\infty}^{\infty} dx \int_{-\infty}^{\infty} dx' 
   \frac{1}{x^2+d^2} \frac{1}{x'^2+d^2}
   {\sum_{m=-\infty}^\infty}{'}
   \frac{(\Delta x)^2+(\Delta t+im\beta)^2/n_B^2}
 {\left[\,(\Delta t+im\beta)^2/n_B^2-(\Delta x)^2\right]^{3}}\;.\nn\\
\end{eqnarray}
Here, the prime means that the $m=0$ term is omitted. With the help of the residue theorem, the integrations above can be directly calculated as
\begin{eqnarray} \label{dt3}
\delta_T ^2
&=&\frac{2 n_B (\chi^{(2)}_x )^2}{\pi^2 n_P^4} 
   \sum_{m=1}^{\infty} 
   \frac{1}{\left(m \beta+2n_P d-2n_B d\right)^2
            \left(m \beta+2n_P d+2n_B d\right)^2}  \nn\\
&&~+\frac{2 n_B \left[ (\chi^{(2)}_y )^2 + (\chi^{(2)}_z )^2 \right]}
        {\pi^2 n_P^4} 
   \sum_{m=1}^{\infty} 
   \frac{4\,d^2(n_B^2+n_P^2)+4\, m\, n_P\, d\,\beta +m^2\beta^2}
        {\left(m \beta+2n_P d-2n_B d\right)^3
         \left(m \beta+2n_P d+2n_B d\right)^3}\;.
\end{eqnarray}

The summation in the result above is hard to find, so in the following we discuss two special cases, i.e. the low temperature ($T \ll d^{-1}$) and high temperature ($T \gg d^{-1}$) limits. As discussed above, to fulfill requirement that the frequency of the background field $\omega_0$ is much smaller than that of the probe field $\omega_1$, the  thermal photon frequency $\beta^{-1}$ should be small compared with $\omega_1$. Therefore, the low and high temperature  limits actually mean $T\ll d^{-1} \ll \omega_1$ and $d^{-1}\ll T \ll \omega_1$ respectively. In the low temperature limit, the flight time fluctuations are found to be in the following form
\begin{eqnarray}\label{result-low}
\delta_T ^2 \approx
\sum_{m=1}^{\infty} 
 \frac{2 n_B \left[ (\chi^{(2)}_x )^2+(\chi^{(2)}_y )^2+(\chi^{(2)}_z )^2 \right]}
      {\pi^2 n_P^4 \beta^4 m^4} 
=\frac{\pi^2 n_B\left[(\chi^{(2)}_x )^2+(\chi^{(2)}_y )^2+(\chi^{(2)}_z )^2 \right] T^4}
      {45 n_P^4 }\;,
\end{eqnarray}
where we have used the relation
\begin{equation}
\sum_{m=1}^{\infty} \frac{1}{m^4}=\frac{\pi^4}{90}\;.
\end{equation}
Therefore, in the low temperature limit, the thermal corrections to the flight time fluctuations is proportional to $T^{4}$, which is  a higher-order correction compared with that induced by vacuum fluctuations proportional to $d^{-4}$, c.f. Eq. (33) in Ref. \cite{analog2}.  

In the high temperature limit, the summation in Eq. (\ref{dt3}) can be approximated by the following integration
\begin{eqnarray}\label{result-high-int}
\delta_T ^2 &\approx&
  \frac{2 n_B(\chi^{(2)}_x )^2}{\pi^2 n_P^4 d^3 \beta}
  \int_{\beta/d}^{\infty} \frac{1}{(x+2n_P-2n_B)^2(x+2n_P+2n_B)^2} dx\nn\\
&&~+\frac{2 n_B \left[ (\chi^{(2)}_y )^2 + (\chi^{(2)}_z )^2\right]}
       {\pi^2 n_P^4 d^3 \beta}
  \int_{\beta/d}^{\infty} 
  \frac{4(n_B^2+n_P^2) + 4 n_P\, x + x^2}{(x+2n_P-2n_B)^3(x+2n_P+2n_B)^3} dx\;.
\end{eqnarray}
Direct calculations show that the leading terms are
\begin{eqnarray}\label{result-high}
\delta_T ^2 &\approx&
\frac{(\chi^{(2)}_x )^2 T}{16 \pi^2 n_B^2 n_P^4 d^3 }
  \left[\frac{2 n_B n_P}{n_P^2-n_B^2} 
 +\ln \left(\frac{n_P-n_B}{n_P+n_B}\right)  \right] \nn\\
&&~+\frac{\left[ (\chi^{(2)}_y )^2 + (\chi^{(2)}_z )^2 \right] T}
        {16 \pi^2 n_B^2 n_P^4 d^3 }
  \left[\frac{n_B n_P (3 n_B^2-n_P^2)}{(n_P^2-n_B^2)^2}
 -\frac{1}{2} \ln \left(\frac{n_P -n_B}{n_P + n_B}\right)\right]\;.
\end{eqnarray}
In this case, the relative flight time variance due to thermal fluctuations is proportional to $d^{-3}T$, which dominates over the contribution of vacuum fluctuations since $T \gg d^{-1}$.

Now a question arises naturally as to whether a very low temperature environment is necessary if we aim at observing the lightcone fluctuations in a nonlinear medium due to vacuum fluctuations. We consider the experiment proposed in Ref. \cite{analog2} with a Cadmium selenide (CdSe) slab. The wavelength of the probe light $\lambda_P=1.06~{\rm \mu m}$, which corresponds to a temperature $T_P=2.15\times10^3~{\rm K}$ if it is taken as the thermal photon wavelength. The thickness of the slab $d$, which determines the frequency regime of the background field  that gives the  dominant contribution, is taken as $d=10.6~{\rm \mu m}$, and the corresponding temperature $T_B=2.15\times10^2~{\rm K}$. Therefore, the room temperature $T\approx3\times10^2~{\rm K}$ is small compared with $T_P$ as required, but it is in neither the low-temperature nor the high-temperature regime compared with $T_B$, so we will calculate Eq. (\ref{dt3}) numerically. Note that the second order susceptibility $\chi^{(2)}_z \approx 1.1\times 10^{-10}~{\rm m/V}$, and the refractive index $n_B=2.43$ at a  wavelength of $10.6~{\rm \mu m}$, and the refractive index $n_P=2.54$ at a wavelength of  $1.06~{\rm \mu m}$  \cite{cdse1,cdse2}. After a unit conversion, the dimensionless ratio $\chi^{(2)}/d^2$ can be written as
\begin{equation}
 \frac{\chi^{(2)}}{d^2}
=6.0\times10^{-8}\left(\frac{\chi^{(2)}}{10^{-12}~{\rm m/V}}\right)
 \left(\frac{1~{\rm \mu m}}{d}\right)^2\;.
\end{equation}
Direct calculations show that the result is $\delta_{\rm rms}=1.8\times10^{-9}$, which is about $18.6\%$ that induced by vacuum fluctuations. Therefore, the main contributions are still from vacuum fluctuations even when the experiment is done at room temperature.

\section{Summary} \label{sec4}

In this paper, we have studied the lightcone fluctuations in a nonlinear medium caused by thermal fluctuations. The effective refractive index for a probe light fluctuates due to the thermal fluctuations of the background electromagnetic fields, which are analogous to the lightcone fluctuations  when gravity is quantized. We have shown that, in the low-temperature limit, the contribution of thermal fluctuations to the flight time fluctuations is proportional to $T^4$, which is negligible compared with that caused by vacuum fluctuations, while in the high-temperature limit, the contribution of thermal fluctuations to the flight time fluctuations increases linearly with $T$, which dominates over that of vacuum fluctuations. Numerical estimations show that, at room temperature, the contributions from thermal fluctuations are still small compared with that from vacuum fluctuations in realistic situations.

\begin{acknowledgments}

This work was supported in part by the NSFC under Grants No. 11435006,  No. 11690034, and No. 11805063.

\end{acknowledgments}


\begin{thebibliography}{00}


\bibitem{Pauli}
W. Pauli, Helv. Phys. Acta Suppl. {\bf 4}, 69 (1956).

\bibitem{Deser}
S. Deser, Rev. Mod. Phys. {\bf 29}, 417 (1957).

\bibitem{DeWitt}
B. S. DeWitt, Phys. Rev. Lett. {\bf 13}, 114 (1964).

\bibitem{Isham71}
C. J. Isham, A. Salam, and J. A. Strathdee, Phys. Rev. D {\bf 3}, 1805 (1971).

\bibitem{Isham72}
C. J. Isham, A. Salam, and J. A. Strathdee, Phys. Rev. D {\bf 5}, 2548 (1972).


\bibitem{Ford95}
L. H. Ford, Phys. Rev. D {\bf 51}, 1692 (1995).

\bibitem{Ford96}
L. H. Ford and N. F. Svaiter, Phys. Rev. D {\bf 54}, 2640 (1996).

\bibitem{Ellis}
J. R. Ellis, N. E. Mavromatos, and D. V. Nanopoulos, Gen. Relativ. Gravit. {\bf 32}, 127 (2000).

\bibitem{Yu03}
H. Yu and P. X. Wu, Phys. Rev. D {\bf 68}, 084019 (2003).

\bibitem{Yu99}
H. Yu and L. H. Ford, Phys. Rev. D {\bf 60}, 084023 (1999).

\bibitem{Yu00}
H. Yu and L. H. Ford, Phys. Lett. B {\bf 496}, 107 (2000).

\bibitem{Yu09}
H. Yu, N. F. Svaiter, and L. H. Ford, Phys. Rev. D {\bf 80}, 124019 (2009).

\bibitem{Mota}
H. F. Mota, E. R. Bezerra de Mello, C. H. G. Bessa, and V. B. Bezerra, Phys. Rev. D {\bf 94}, 024039 (2016).





%%%%%%%%%%%%%%%%%%%%%%%%


\bibitem{analog1}
L. H. Ford, V. A. De Lorenci, G. Menezesc, and N. F. Svaiter, Ann. Phys. (Amsterdam) {\bf 329},  80 (2013).

\bibitem{analog2}
C. H. G. Bessa, V. A. De Lorenci, L. H. Ford, and N. F. Svaiter, Ann. Phys. (Amsterdam) {\bf 361},  293 (2015).

\bibitem{analog3}
C. H. G. Bessa, V. A. De Lorenci, and L. H. Ford, Phys. Rev. D {\bf 90}, 024036 (2014).

\bibitem{analog4}
C. H. G. Bessa, V. A. De Lorenci, L. H. Ford, and C. C. H. Ribeiro, Phys. Rev. D {\bf 93}, 064067 (2016).



\bibitem{no-book}
R. W. Boyd, {\it Nonlinear Optics}, 3rd ed. (Academic Press, New York, 2008).


%%%%%%%%%%%%%%%%%%%%%%

\bibitem{qftcs}
N. D. Birrell and P. C. W. Davies, {\it Quantum Fields in Curved Space} (Cambridge University Press, Cambridge, England, 1982).

\bibitem{Glauber}
R. J. Glauber and  M. Lewenstein, Phys. Rev. A {\bf 43}, 467 (1991).

\bibitem{Barnett}
S. M. Barnett, B. Huttner, and  R. Loudon, Phys. Rev. Lett. {\bf 68}, 3698 (1992).

\bibitem{cdse1}
R. L. Herbst and R. L. Byer, Appl. Phys. Lett. {\bf 19}, 527 (1971).

\bibitem{cdse2}
G. C. Bhar, Appl. Opt. {\bf 15}, 305 (1976).



\end{thebibliography}
\end{document}